\documentstyle[12pt,aasms4]{article}

\begin{document}

\title{An Infrared Determination of the Reddening 
and Distance to Dwingeloo 1}
\author{Valentin D. Ivanov, Almudena Alonso-Herrero, Marcia J. Rieke and 
	Don McCarthy}
\affil{Steward Observatory, The University of Arizona, 933 N. Cherry 
Ave, Tucson, AZ 85721}
\author{E-mail: vdivanov@as.arizona.edu, aalonso@as.arizona.edu, 
mrieke@as.arizona.edu, mccarthy@as.arizona.edu}

\begin{abstract}
We present for the first time infrared observations of the nearby 
highly obscured galaxy Dwingeloo 1 (Dw1), including deep $H$-band 
imaging covering a total of $4.9\arcmin \times 4.9\arcmin$, together with 
$J$ and $K_{\rm s}$ imaging of the central $2.5\arcmin \times 2.5\arcmin$. 
We used the small dispersion of the intrinsic infrared colors of spiral 
galaxies to determine an infrared $H$-band extinction of 
$A_H = 0.47\pm0.11$\,mag towards Dw1. In using infrared colors, the 
uncertainties in the reddening and distance are reduced by a factor of 
three. The $H$-band magnitude corrected for extinction and the 
infrared Tully-Fisher relation are then used to estimate a  
distance modulus of $(m-M)_0 = 28.62\pm0.27$, and thus a distance 
of $d = 5.3^{+0.7}_{-0.6}\,$Mpc,
which puts Dw1 at the far end of the IC~342/Maffei 1 \& 2 group. Our 
result is largely independent of the nature of the reddening law 
because we estimated both the reddening and the distance at the same 
wavelength range. 
\end{abstract}


\section{Introduction}

Dwingeloo 1 (Dw1) is a large SBb/c galaxy, discovered both in a systematic 
H\,{\sc i} emission survey
of the northern part of the Milky Way in search of obscured galaxies in 
the Zone of Avoidance by Kraan-Korteweg et al. \markcite{kraa94}(1994),
and independently by Huchtmeier et al. \markcite{huch95}(1995). The 
knowledge of the local mass distribution has implications for the 
peculiar velocity field, the direction and amplitude of the Local 
Group acceleration,  the determination of parameters such as $\Omega_0$
and $H_0$, and 
on the understanding of the formation and evolution of groups of 
galaxies (e.g., \markcite{peeb94}Peebles 1994;
\markcite{mari98}Marinoni et al. 1998). The discovery of this galaxy proved 
a long standing suspicion that the tidal disruptions of Maffei 2 
may be due to the presence of another massive galaxy nearby
\markcite{hurt93}(e.g., Hurt et al. 1993).

Dw1 lies in the direction of the IC~342/Maffei 1 \& 2 group 
of galaxies, about 2 degrees away 
from Maffei 2. This corresponds to a physical separation of 
175\,kpc assuming that Dw1 and  Maffei 2  are at a distance of 
5\,Mpc. Being the nearest barred spiral system, 
Dw1 offers a unique possibility to study the effect of the bar at 
high spatial resolution. The discoverers classified Dw1 as an SBb or 
SBc galaxy (T=4) and measured  an angular diameter of 4.2\,arcmin. Later on, 
\markcite{mcca97}McCall \& Buta (1997) re-classified it to SB(s)cd 
with an angular diameter of 9.9\,arcmin at $\mu_I = 25.0\,$mag arcsec$^{-2}$ 
based on deep optical $I$-band imaging. 
\markcite{burt96}Burton et al. (1996) extensively studied the 
neutral hydrogen content of Dw1 and measured  H\,{\sc i} 
profile widths at 20\% and 50\% level of $201.2\pm0.4\,$km s$^{-1}$ and 
$187.6\pm0.6\,$km s$^{-1}$ respectively. The measured inclination of 
the gaseous disk was $51\pm2\,$degrees and the position angle 112\,degrees, 
with the major axis aligned with the bar. 

Since the discovery of Dw1 the determination of its distance
has been hampered by the poorly known Galactic extinction. 
Optical $VRIH_{\alpha}$ imaging, long slit spectroscopy, and {\it IRAS}
observations were summarized by \markcite{loan96}Loan et al. (1996), who 
used a number of methods to estimate the foreground extinction 
towards Dw1. The optical color excesses yielded $A_V=7.8\pm3.0\,$,mag, the 
measured Galactic H\,{\sc i} column density  $A_V=4.5\,$mag, and 
the $100\,\mu$m {\it IRAS} flux  $A_V=3.2\,$mag. Finally, 
they applied optical $I$ and $R$-band 
Tully-Fisher relations to obtain distances ranging from 1.3 to 6.7\,Mpc, 
with an average value of about 4\,Mpc (assuming $H_0 = 75\,$km s$^{-1}$ 
Mpc$^{-1}$). Their main source of uncertainty was the value of the 
Galactic extinction. 
\markcite{phil97}Phillipps \& Davies (1997) challenged these extreme 
distance estimates on the basis of the very narrow span of central 
surface brightness in present day spiral galaxies. Their best 
estimate for the extinction ($A_B=6\,$mag) places Dw1 at a distance of 
$3.1-3.6\,$Mpc. These authors also employed the diameter version 
of the Tully-Fisher relation (Persic, Salucci, \&
Stel 1996) and  obtained a distance of 2.7\,Mpc. 

The primary goal of this study was to put stronger constraints on the 
foreground extinction, and to obtain a better estimate of the distance 
to Dw1. We chose to use infrared colors 
because of their small intrinsic variations among  spiral 
galaxies \markcite{aaro77}(Aaronson 1977). In addition, 
the extinction in the $H$-band 
is about three times smaller than in the $I$-band and about six times 
smaller than in the $V$-band \markcite{riek85}(Rieke \& Lebofsky 1985). 
Finally, the infrared Tully-Fisher (IRTF) relation shows a 
smaller intrinsic scatter than its optical counterpart 
(\markcite{aaro79}Aaronson, Huchra, \&
Mould  1979; Freedman 1990; \markcite{pele93}Peletier 
\& Willner 1993). The IRTF relation  allows us to determine 
both the extinction and the 
distance at the same wavelength range, minimizing the errors arising 
from possible variations of the reddening law.

\section{Observations and Data reduction}

We obtained $JHK_{\rm s}$ imaging of Dw1 using a $256\times256$ 
NICMOS3 array at the 2.3-m Bok Telescope of the University of Arizona 
on Kitt Peak, with a plate scale of $0.6\,$arcsec pixel$^{-1}$
during a number of observing runs. We constructed a deep 
$4.9\arcmin \times 4.9\arcmin$ $H$-band mosaic of Dw1, whereas the 
$J$ and $K_{\rm s}$ images only covered the central $2.5\arcmin
\times 2.5\arcmin$.  Additional $H$-band imaging   
using a $1024\times 1024$ array and  plate scale $0.5\,$arcsec 
pixel$^{-1}$ was 
obtained at the same telescope on a subsequent observing run 
to calibrate the deep $H$-band imaging.
The observational strategy consisted of taking galaxy images interleaved 
with sky images $6\arcmin-7\arcmin$ away from Dw1. 
Details of the observations are listed in Table~1. 

The data reduction included subtraction of dark current frames, 
flat-fielding with median combined empty sky frames, and 
sky subtraction. The mosaics were 
constructed by shifting the images to a common position with 
cubic spline interpolation. The photometric calibration was 
performed using observations of standard stars from the lists of 
\markcite{elia82}Elias et al. (1982) and 
\markcite{hunt98}Hunt et al. (1998) when conditions were photometric.  
Conditions were non-photometric for the $H$-band 
mosaic, and this image was self-calibrated using the $1024\times1024$
data. In the next section we will be making use of infrared colors of 
spiral galaxies to derive an estimate of the extinction to Dw1. The 
colors for spiral galaxies were obtained with the $K$-band filter, 
whereas our measurements were taken with the $K_{\rm s}$ (short-$K$)
filter. Therefore it is necessary to determine what the difference
is between the two filters.
Very recently Persson et al. (1998) have obtained a new set of 
$JHKK_{\rm s}$ photometric standards. In their study the average 
difference between the 
$K$ and $K_{\rm s}$ magnitudes  (for the red standards) is 0.0096\,mag, with a 
standard deviation of 0.017\,mag. We assume that the use of the
$K_{\rm s}$ filter instead of the $K$ filter introduces an extra 
uncertainty in the photometric calibration of this filter of $\pm 0.02\,$mag. 
The errors associated with the photometric calibration are
0.05, 0.07, 0.06 mag in $J$, $H$ and $K_{\rm s}$ respectively.

In Figure~1 we display the $H$-band mosaic on a logarithmic scale. 
As can be seen from this figure, there is a large number of foreground stars 
which need to be removed prior to analyzing the data. The star removal 
from the $JHK_{\rm s}$ images was done interactively. The affected pixels 
were then replaced with a 
linear surface fit to a circular annulus around each star. The 
automatic procedures failed largely because of PSF variations 
under non-photometric conditions. A bright star located southwest of the 
galaxy posed a major problem, and was masked out throughout the 
data reduction. We assumed radial symmetry  
and replaced the region within about 90 arcsec from the star with 
the data from the opposite side of the galaxy. We performed the 
photometry on the {\it cleaned} images measuring the flux within elliptical 
isophotes with fixed position angle and ellipticity as determined 
from H\,{\sc i} observations (\markcite{burt96}Burton et al. 1996). 

\section{Discussion}

\subsection{Colors and Extinction}

The surface brightness profiles in $JHK_{\rm s}$ are 
given in Table~2, and displayed in Figure~2 along with  
the radial distribution of the $J-H$, $H-K_{\rm s}$ and 
$J-K_{\rm s}$ colors, and the total apparent magnitudes in $JHK_{\rm s}$. 
In both Table~2 and Figure~2 the error bars 
represent the combined $3 \sigma$ 
variations from photon statistics, sky background variations, the 
elliptical isophotal fitting, and the photometric calibration. 
From the radial distribution of the colors, it is clear that the 
center of the galaxy appears slightly 
redder than the outer regions, as it is the case with the 
infrared colors of most spiral galaxies 
(\markcite{tern94}Terndrup et al. 1994\markcite{dejo96}; de Jong 
1996). \markcite{loan96}Loan et al. (1996) report inverse 
optical color gradients in Dw1, however, this may be the 
result of active star formation along the bar as found in some barred spirals 
(\markcite{shaw95}Shaw et al. 1995).

Before we tried to obtain an estimate of the extinction using the 
infrared colors, we fitted straight lines to assess the variation of the 
colors with increasing apertures,  

\begin{equation}
H-K_{\rm s} = 0.435(\pm0.042)-0.072(\pm0.071) R\\
\end{equation}

\begin{equation}
J-H = 1.099(\pm0.040)-0.056(\pm0.067) R\\
\end{equation}

\begin{equation}
J-K_{\rm s} = 1.532(\pm0.042)-0.125(\pm0.071) R\\
\end{equation}

where $R$ is the semi-major axis in units of arcmin. These color 
gradients imply a change of $0.06-0.13\,$mag within the inner 
2\,arcmin of Dw1, which may cause  
significant uncertainties in the reddening estimate. The smaller field of 
view of the $J$ and $K_{\rm s}$ images  
prevents us from obtaining the total colors of Dw1.  However, 
the total flux in the $JHK_{\rm s}$ bands is dominated by the inner 
$2\arcmin$-diameter region (see Figure~2, bottom panel). Hence, we 
adopt the total observed colors at radial distance of 1 arcmin from the center
of the galaxy to be representative for the whole galaxy 
($J-H = 1.04\pm0.10\,$mag, $J-K_{\rm s} = 1.40\pm0.11\,$mag, 
$H-K_{\rm s} = 0.36\pm0.11\,$mag). To estimate the color excesses 
we compared these colors to the mean 
integrated colors of SBb-SBcd galaxies: 
$J-H = 0.73\pm0.02\,$mag, $J-K_{\rm s} = 0.94\pm0.03\,$mag, 
$H-K_{\rm s} = 0.21\pm0.02\,$mag 
(\markcite{aaro77}Aaronson 1977). 
We have increased the errors 
in Aaronson's colors to account for the uncertain Hubble type of Dw1. 
The color excesses  were converted into $H$-band ($A_H$), visual 
($A_V$) and $B$-band ($A_B$) extinctions using Rieke \&
Lebofsky (1985, RL85) and Mathis (1990, M90) extinction laws. 
The results are summarized in Table~3. 
Henceforth we will use $A_H = 0.47\pm0.11\,$mag from  
Mathis (1990) extinction law for the sake of compatibility 
with previous work. This value 
is close to the estimate based on the {\it IRAS} $100\,\mu$m flux
(\markcite{loan96}Loan et al. 1996).

Finally, we used the combined optical - near infrared colors to verify 
our result. \markcite{loan96}Loan et al. (1996) reported the following 
total apparent (not corrected for reddening) magnitudes for Dw 1: 
$m_I = 10.7\pm0.2\,$mag, $m_R = 12.2\pm0.2\,$mag, and
$m_V = 14.0\pm0.5\,$mag. We measured a total
$H$-band magnitude  $m_H = 8.3\pm0.2\,$mag, and 
compared the observed colors with the intrinsic colors as determined 
by \markcite{dejo96}de Jong (1996): $I-H = 1.44\pm0.20\,$mag, 
$R-H = 2.01\pm0.20\,$mag, and $V-H = 2.50\pm0.20\,$mag. 
\markcite{math90}Mathis (1990) extinction law yields 
$A_H = 0.56\pm0.23\,$mag, $0.58\pm0.12\,$mag, and $0.47\pm0.11\,$mag 
respectively, in good agreement with our infrared estimates.

\subsection{Tully-Fisher Distance to Dw1} 

In order to determine the distance to Dw1, we chose 
to apply the IRTF relation because of its lower
intrinsic dispersion and the reduced extinction in the $H$-band, with 
the additional advantage that the extinction and the distance are 
estimated at the same wavelength. The IRTF relation was 
pioneered by \markcite{aaro79}Aaronson et al. (1979). However, we chose to 
employ the IRTF relation calibrated by \markcite{free90}Freedman (1990) using 
local galaxies with Cepheid based distances, and the relations of 
\markcite{pele93}Peletier \& Willner (1993) calibrated relative to 
the distance of the Ursa Major galaxy cluster. 

\markcite{free90}Freedman's (1990) calibration for the IRTF relation leads
to the following expression:  
$H^{\rm abs}_{-0.5} = -10.26(\pm0.49)(\log{\Delta}V_{20}(0)-2.5)-21.02
(\pm0.08)$,
where ${\Delta}V_{20}(0)$ is the inclination corrected 
$20\%$ level H\,{\sc i} velocity profile width in km s$^{-1}$. 
$H^{\rm abs}_{-0.5}$ is the absolute $H$-band magnitude within a 
circular aperture with diameter $A$, for which  $\log(A/D_0)=-0.5$, with $D_0$ 
being the $B$-band isophotal diameter at $\mu_{B,0} = 25\,$mag arcsec$^{-2}$.
Using ${\Delta}V_{20}(0) = 259.0\pm0.5\,$km s$^{-1}$, which is 
\markcite{burt96}Burton et al.'s (1996) value corrected for the 
inclination of the galaxy, the above expression predicts an absolute 
$H$-band magnitude  $H^{\rm abs}_{-0.5} = -20.13\pm0.21\,$mag.

Although $B$-band surface photometry for Dw1 is not available, we can 
make use of the intrinsic integrated colors for Sb-Sc galaxies 
$B-H = 3.28\pm0.14$ (\markcite{dejo96}de Jong 1996)
to estimate $D_0$. The   
$\mu_{B,0} = 25\,$mag arcsec$^{-2}$ isophote corresponds to  
an $H$-band surface brightness (corrected for extinction) of  
$\mu_{H,0} = 21.72\pm0.14\,$mag arcsec$^{-2}$. Taking into account the 
$H$-band extinction this is equivalent to an observed (not corrected
for extinction) value of $\mu_{H} = 22.19\pm0.17\,$mag arcsec$^{-2}$. As
can be seen from Figure~2 (upper panel), this value exceeds 
the boundaries of the $H$-band mosaic. However, the surface brightness
profile can be easily extrapolated. We fitted an exponential disk 
($\mu_H\propto e^{-r/r_d}$, where $r$ is the 
semi-major axis and $r_d$ is the disk scale length) to the surface 
brightness profile from a radial distance of 2\,arcmin  
outwards where the bulge contribution in negligible, and estimated 
a value of $D_0 = 8.5\pm0.8$ arcmin. The apparent $H$-band magnitude (not 
corrected for extinction) for a circular aperture  with diameter of  
$A_{-0.5}=2.7\pm0.2\,$arcmin 
is then $m(H^{\rm app}_{-0.5}) = 8.96\pm0.12\,$mag, which provides  
a distance modulus of $(m-M)_0 = 28.62\pm0.26$ and a distance of 
$d = 5.3^{+0.7}_{-0.6}\,$Mpc. 

The IRTF relation was initially calibrated for circular apertures  
because only single-pixel  detectors 
were used at the time (\markcite{aaro79}Aaronson, 
Huchra, \& Mould 1979). 
Naturally, one would expect a transition to elliptical apertures to 
reduce the internal dispersion of the IRTF relation because they 
correct for the galaxy inclination. \markcite{pele93}Peletier \&
Willner (1993) 
studied the problem in detail and 
reported no significant 
change in the calibration of the IRTF relation when elliptical 
apertures were used. A possible 
explanation is that the higher internal absorption, which increases
with inclination, may cancel out the projection effect. 
\markcite{pele93}Peletier \& Willner (1993) used elliptical apertures 
and obtained the following calibration for spiral galaxies in the Ursa 
Major galaxy cluster: 
$\log {\Delta}V_{20}(0) = 
-0.085 (H^{\rm abs,e}_{-0.5} + 30.95 - 9.0) + 2.603$. We used 
a distance modulus to 
Ursa Major Cluster of $(m-M)_0 = 30.95\pm0.17\,$mag 
(\markcite{peir88}Pierce \& Tully 1988).
This gives $H^{\rm abs,e}_{-0.5} = -19.72\pm0.52\,$mag. 
However, we are still left with 
the problem of extrapolating the observed total luminosity profile 
out to $D_0$. The measured total (not corrected for extinction) $H$-band 
magnitude within an elliptical aperture with major axis 
$2.7\pm0.2\,$arcmin and axial ratio 1.56 
(\markcite{loan96}Loan et al. 1996) is 
$m(H^{\rm app}_{-0.5}) = 9.19\pm0.43\,$mag. 
Correcting for the reddening, we obtained a distance modulus of  
$(m-M)_0 = 28.44\pm0.69\,$mag and 
distance $d = 4.9^{+1.8}_{-1.3}\,$Mpc. 
The intrinsic spread of \markcite{pele93}Peletier \& Willner (1993) 
IRTF relation and the uncertain distance to the Ursa Major Cluster 
account for the increased errors of this estimate.

We can now use our reddening estimate to correct the optical $I$ 
and $R$-band photometry of \markcite{loan96}Loan et al. (1996), 
and use the optical Tully-Fisher relations to obtain another  
distance estimate.  The two reddening laws discussed in the preceding 
section predict the same $A_I$/$A_H$ ratio within a few percent. 
Adopting \markcite{math90}Mathis (1990) extinction law, we find 
$I$- and $B$-band extinctions of 
$A_I = 1.3\pm0.3\,$mag, and $A_B = 3.6\pm0.8\,$mag lower than  
$A_B = 4.3\,$mag in \markcite{loan96}Loan et al. (1996). This puts 
Dw1 at an average distance of $5.5^{+0.8}_{-0.7}$ Mpc.

These distance determinations place Dw1 behind NGC~1560 at 
$3.5\pm0.7\,$Mpc, UGCA~105 at $3.8\pm0.9\,$Mpc 
(\markcite{kris95}Krismer, Tully, \& Gioia 1995), and Maffei 1 at 
$4.2\pm0.5\,$Mpc (\markcite{lupp93}Luppino \& Tonry 1993). 

\section{Conclusions} 

We have obtained deep near-infrared imaging of the highly obscured 
galaxy Dw1. The observed infrared colors were used  to obtain a very 
accurate estimate of the extinction in the $H$-band, 
$A_H=0.47\pm0.11\,$mag. This value was confirmed by the optical - 
near infrared color excesses, and is close to the estimate based on 
the {\it IRAS} $100\,\mu$m flux (\markcite{loan96}Loan et al. 1996).
Our approach is more reliable than previous works in that we did not 
make any additional assumptions for the relation between $A_V$ and 
H\,{\sc i}, or the {\it IRAS} $100\,\mu$m emission. In addition, the IRTF 
relation allowed us to estimate both the reddening and distance at the 
same wavelength range. This makes our results largely independent of 
the choice of the reddening law, with the additional advantage that 
the IRTF shows a smaller dispersion than its optical counterpart.
Finally, the infrared reddening estimates are more reliable than those 
in the optical because infrared colors of spiral galaxies 
show a smaller intrinsic dispersion, 
and are less sensitive to the history of star formation than 
optical colors (\markcite{vazd96}Vazdekis et al. 1996).

The IRTF relation (Freedman 1990) yielded a 
distance of $d = 5.3^{+0.7}_{-0.6}\,$Mpc 
which places Dw1 at the far end of the IC~342/Maffei~1 \& 2 group. 
We also confirmed that Dw1 has an angular diameter greater than 7 arcmin, 
larger than the originally measured value of 4.2 arcmin. 

\vspace{2cm}

During the course of this work VDI and AA-H were supported by the 
National Aeronautics and Space Administration on grant NAG 5-3042 
through the University of Arizona.  The $256 \times 256$ camera 
was supported by NSF Grant AST-9529190. We are 
grateful to the anonymous referee for comments which helped improve the 
paper.

\newpage

{\bf Figure Captions}

Figure~1.--- $H$-band mosaic of Dw1 displayed on a logarithmic 
scale. The orientation is north up, east to the left. The field
of view is $4.7\arcmin \times 4.9\arcmin$.

Figure~2.--- {\it Upper panel:} Observed surface brightness profiles in 
$J$, $H$ and $K_{\rm s}$ as a function of the semi-major axis. The 
dashed line is our exponential disk fit (see text).
{\it Middle panel:} Radial distribution of the $J-H$, $H-K_{\rm s}$ 
and $J-K_{\rm s}$ colors. The straight lines represent a linear fit 
to the color gradients as a function of the semi-major axis (see text). 
{\it Bottom panel:} Observed total magnitude as a function of the 
semi-major axis in $J$, $H$ and $K_{\rm s}$. 
In all three panels, the vertical bars represent $3\sigma$ errors.

\newpage

\begin{deluxetable}{llrrrc}
\tablenum{1}
\tablewidth{0pt}
\tablecaption{Log of the observations.}
\tablehead{
\colhead{Detector} &\colhead{Filter} & \colhead{$t_{\rm exp}$} &
\colhead{Date} & \colhead{Conditions} & \colhead{seeing} \nl
               &           & \colhead{(s)} & & & \colhead{(arcsec)}}
\startdata
$256\times256$   & $J$         &  600 &  Nov 5 1998 & photometric     & 1.2 \nl
                 & $J$         &  780 &  Jan 1 1999 & non-photometric & 1.5 \nl
                 & $H$         & 1800 & Dec 30 1998 & non-photometric & 1.1 \nl
                 & $K_{\rm s}$ &  720 &  Nov 5 1998 & photometric     & 1.1 \nl
$1024\times1024$ & $H$         &  760 &  Jan 6 1999 & photometric     & 1.0 \nl
\enddata
\end{deluxetable}

\newpage

\begin{deluxetable}{lrrrrrr}
\tablenum{2}
\tablewidth{0pt}
\tablecaption{Surface brightness in $JHK_{\sc s}$ for Dw1.}
\tablehead{
\colhead{Semi-major}       & 
\colhead{$\mu_J$}         & \colhead{$\sigma(\mu_J)$} & 
\colhead{$\mu_H$}         & \colhead{$\sigma(\mu_H)$} & 
\colhead{$\mu_K$}         & \colhead{$\sigma(\mu_K)$}\nl
\colhead{(1)} & \colhead{(2)} &\colhead{(3)} &\colhead{(4)} &
\colhead{(5)} &\colhead{(6)} &\colhead{(7)}}
\startdata
1.2 & 18.35 & 0.14 & 17.21 & 0.15 & 16.80 & 0.18 \nl 
3.6 & 18.87 & 0.06 & 17.88 & 0.06 & 17.42 & 0.07 \nl 
6.0 & 19.09 & 0.07 & 18.05 & 0.06 & 17.63 & 0.07 \nl 
8.4 & 19.29 & 0.07 & 18.18 & 0.06 & 17.74 & 0.07 \nl 
10.8 & 19.39 & 0.07 & 18.36 & 0.06 & 17.88 & 0.07 \nl 
13.2 & 19.57 & 0.06 & 18.50 & 0.06 & 18.07 & 0.07 \nl 
15.6 & 19.72 & 0.06 & 18.62 & 0.07 & 18.20 & 0.07 \nl 
18.0 & 19.88 & 0.06 & 18.74 & 0.07 & 18.33 & 0.07 \nl 
20.4 & 19.98 & 0.06 & 18.87 & 0.07 & 18.47 & 0.07 \nl 
22.8 & 20.08 & 0.06 & 18.96 & 0.07 & 18.56 & 0.07 \nl 
25.2 & 20.18 & 0.06 & 19.07 & 0.07 & 18.69 & 0.07 \nl 
27.6 & 20.27 & 0.06 & 19.15 & 0.07 & 18.78 & 0.07 \nl 
30.0 & 20.36 & 0.06 & 19.25 & 0.07 & 18.88 & 0.07 \nl 
32.4 & 20.40 & 0.06 & 19.29 & 0.07 & 18.91 & 0.07 \nl 
34.8 & 20.46 & 0.06 & 19.34 & 0.07 & 19.00 & 0.07 \nl 
37.2 & 20.50 & 0.06 & 19.41 & 0.07 & 19.06 & 0.07 \nl 
39.6 & 20.55 & 0.06 & 19.49 & 0.06 & 19.09 & 0.07 \nl 
42.0 & 20.59 & 0.06 & 19.55 & 0.06 & 19.15 & 0.07 \nl 
44.4 & 20.64 & 0.06 & 19.59 & 0.06 & 19.19 & 0.07 \nl 
46.8 & 20.70 & 0.06 & 19.62 & 0.06 & 19.27 & 0.07 \nl 
49.2 & 20.74 & 0.06 & 19.69 & 0.06 & 19.33 & 0.07 \nl 
51.6 & 20.76 & 0.06 & 19.76 & 0.06 & 19.38 & 0.07 \nl 
54.0 & 20.80 & 0.06 & 19.79 & 0.06 & 19.43 & 0.07 \nl 
56.4 & 20.87 & 0.06 & 19.83 & 0.06 & 19.43 & 0.07 \nl
58.8 & 20.91 & 0.06 & 19.89 & 0.06 & 19.47 & 0.07 \nl
\hline
61.2 &       &      & 19.91 & 0.06 &       &      \nl
70.8 &       &      & 20.04 & 0.06 &       &      \nl
80.4 &       &      & 20.13 & 0.06 &       &      \nl
90.0 &       &      & 20.21 & 0.06 &       &      \nl
99.6 &       &      & 20.35 & 0.06 &       &      \nl
109.2 &       &      & 20.44 & 0.07 &       &      \nl
118.8 &       &      & 20.53 & 0.07 &       &      \nl
128.4 &       &      & 20.68 & 0.07 &       &      \nl
138.0 &       &      & 20.85 & 0.07 &       &      \nl
147.6 &       &      & 21.01 & 0.08 &       &      \nl
157.2 &       &      & 21.13 & 0.08 &       &      \nl
166.8 &       &      & 21.08 & 0.21 &       &      \nl
176.4 &       &      & 21.38 & 0.10 &       &      \nl
\enddata
\tablecomments{Column~(1) is the semimajor axis in arcsec. 
Columns~(2), (4) and (6) are the observed surface brightnesses
expressed in mag arcsec$^{-2}$ in $J$, $H$, and $K_{\rm s}$. 
Column~(3), (5) and (7) are the 3$\sigma$ errors (see text).} 
\end{deluxetable}

\newpage

\begin{deluxetable}{llll}
\tablenum{3}
\tablewidth{0pt}
\tablecaption{Colors and redenning.}
\tablehead{
\colhead{} &\colhead{$(J-H)$} & \colhead{$(J-K_s)$} & \colhead{$(H-K_s)$}}
\startdata
Colors\tablenotemark{a}  & 
     $1.04$ & $1.40$ & $0.36$ \nl
Photometric errors  &
     $0.09$ & $0.08$ & $0.09$ \nl
Fit errors \tablenotemark  & 
     $0.08$ & $0.08$ & $0.08$ \nl
Total errors  &
     $0.12$ & $0.11$ & $0.12$ \nl
Intrinsic colors &
     $0.73\pm0.02$ & $0.94\pm0.03$ & $0.21\pm0.02$ \nl
Color excesses &
     $0.31\pm0.12$ & $0.46\pm0.11$ & $0.15\pm0.12$ \nl
$A_H$ (M90) & $0.51\pm0.20$ & $0.47\pm0.11$ &
     $0.39\pm0.31$ \nl
$A_H$ (RL85) & $0.51\pm0.20$ & $0.46\pm0.11$ &
     $0.42\pm0.33$ \nl
$A_V$ (M90) & $2.93\pm1.13$ & $2.70\pm0.63$ &
     $2.20\pm1.76$ \nl
$A_V$ (RL85) & $2.90\pm1.12$ & $2.76\pm0.64$ &
     $2.38\pm1.90$ \nl
$A_B$ (M90)& $3.88\pm1.50$ & $3.57\pm0.83$ &
     $2.91\pm2.33$ \nl
$A_B$ (RL85) & $3.84\pm1.48$ & $3.65\pm0.85$ &
     $3.15\pm2.52$ \nl
\enddata
\tablenotetext{a}{Colors (not corrected for extinction) at 1 arcsec radial 
distance as obtained from the fits (see text).}
\end{deluxetable}

\newpage

\begin{figure}
\figurenum{1}
\plotfiddle{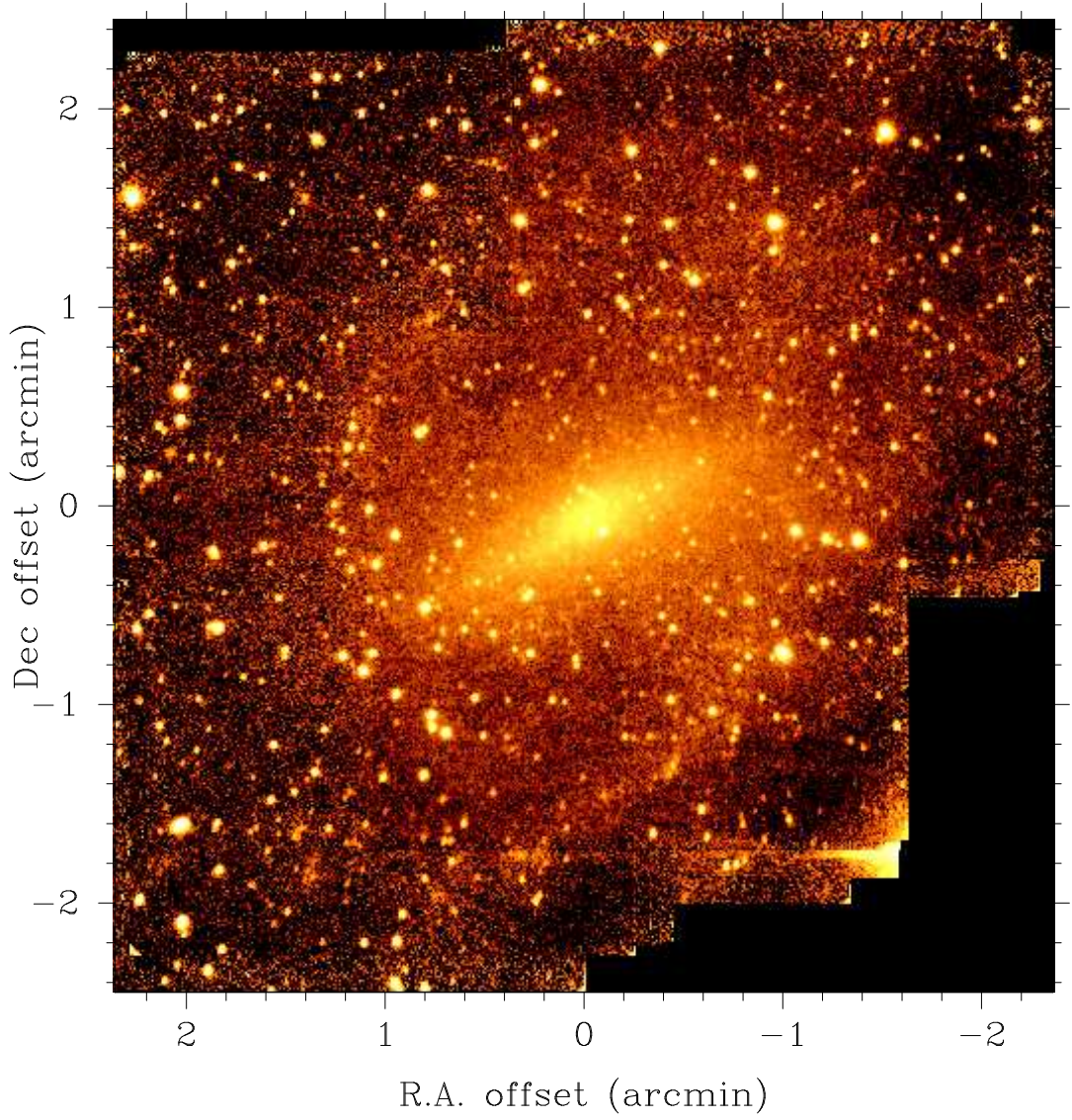}{425pt}{0}{80}{80}{-270}{-80}
\caption{$H$-band mosaic of Dw1 displayed on a logarithmic 
scale. The orientation is north up, east to the left. The field
of view is $4.7\arcmin \times 4.9\arcmin$.}
\end{figure}

\begin{figure}
\figurenum{2}

\vspace{3cm}

\plotfiddle{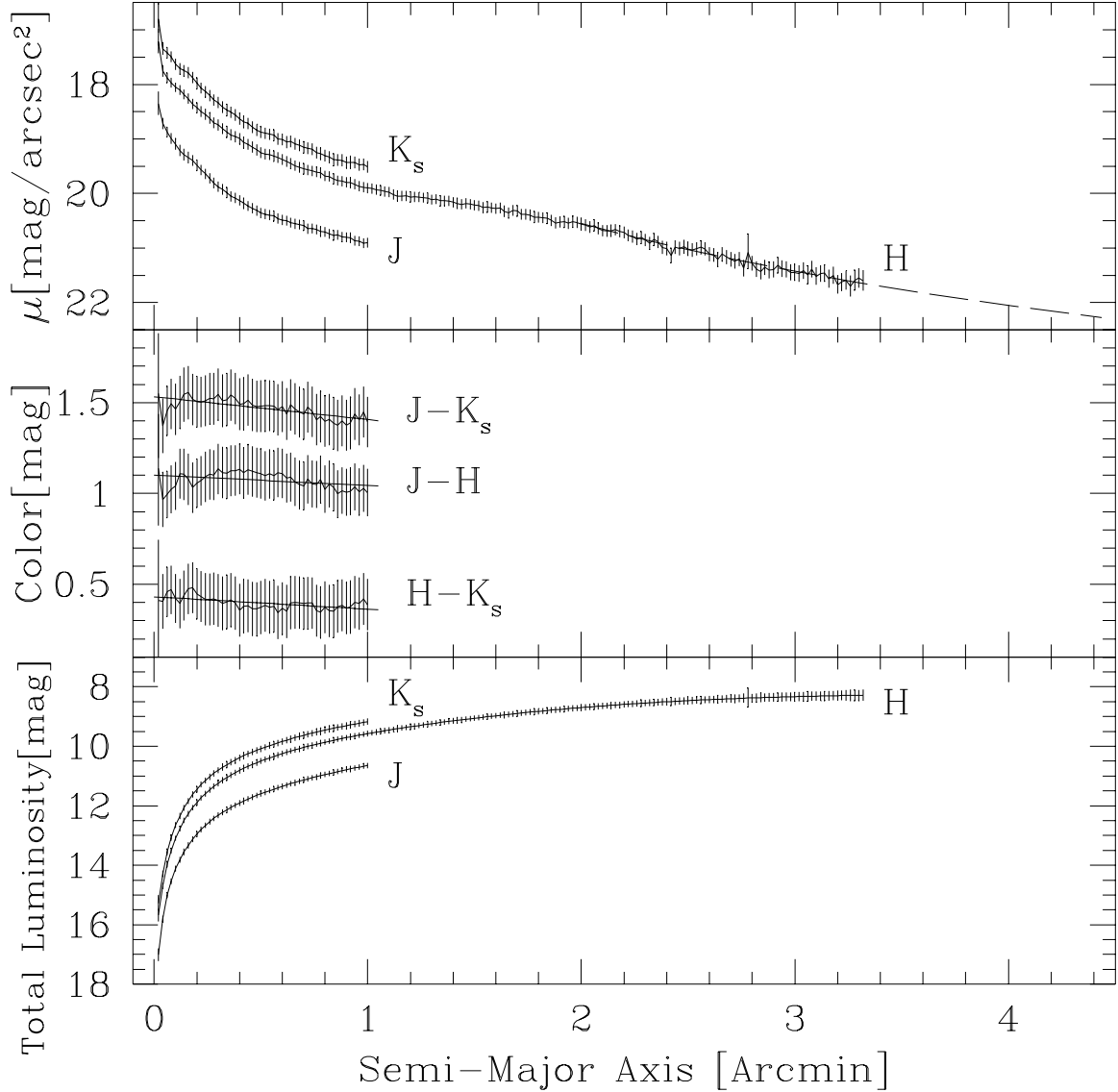}{425pt}{0}{80}{80}{-250}{-80}
\caption{{\it Upper panel:} Observed surface brightness profiles in 
$J$, $H$ and $K_{\rm s}$ as a function of the semi-major axis. The 
dashed line is our exponential disk fit (see text).
{\it Middle panel:} Radial distribution of the $J-H$, $H-K_{\rm s}$ 
and $J-K_{\rm s}$ colors. The straight lines represent a linear fit 
to the color gradients as a function of the semi-major axis (see text). 
{\it Bottom panel:} Observed total magnitude as a function of the 
semi-major axis in $J$, $H$ and $K_{\rm s}$. 
In all three panels, the vertical bars represent $3\sigma$ errors.}
\end{figure}

\end{document}